\documentclass[twocolumn,floatfix,amsmath,amssymb,prb,showpacs]{revtex4-1}
\usepackage{graphicx}
\usepackage{dcolumn}
\usepackage{bm}
\usepackage{amsmath}
\begin{document}
\title{Conductance anomaly near the Lifshitz transition in strained bilayer graphene}
\author{Diana A. Gradinar}
\email[Electronic mail: ]{d.cosma@lancaster.ac.uk}
\affiliation{Department of Physics, Lancaster University, LA1 4YB Lancaster, United Kingdom}
\author{Henning  Schomerus}
\affiliation{Department of Physics, Lancaster University, LA1 4YB Lancaster, United Kingdom}
\author{Vladimir I. Fal'ko}
\affiliation{Department of Physics, Lancaster University, LA1 4YB Lancaster, United Kingdom}
\date{\today}
\begin{abstract}
Strain qualitatively changes the low-energy band structure of bilayer graphene, leading to the appearance of a pair of low-energy Dirac cones near each corner of the Brillouin zone, and a Lifshitz transition,~(a saddle point in the dispersion relation) at an energy proportional to the strain~[M.~Mucha-Kruczynski, I.L.~Aleiner, and V.I.~Fal'ko, Phys. Rev. B $\bf{84}$, 041404 (2011)]. Here, we show that in the vicinity of the Lifshitz transition the conductance of a ballistic n-p and n-p-n junction exhibits an anomaly: a non-monotonic temperature and chemical potential dependence, with the size depending on the crystallographic orientation of the principal axis of the strain tensor. This effect is characteristic for junctions between regions of different polarity ($n$-$p$ and $n$-$p$-$n$ junctions), while there is no anomaly in junctions between regions of the same polarity ($n$-$n^{\prime}$ and $n$-$n^{\prime}$-$n$ junctions).
\end{abstract}


\maketitle

\section{Introduction}
\begin{figure}[t]
\includegraphics[width=1\columnwidth]{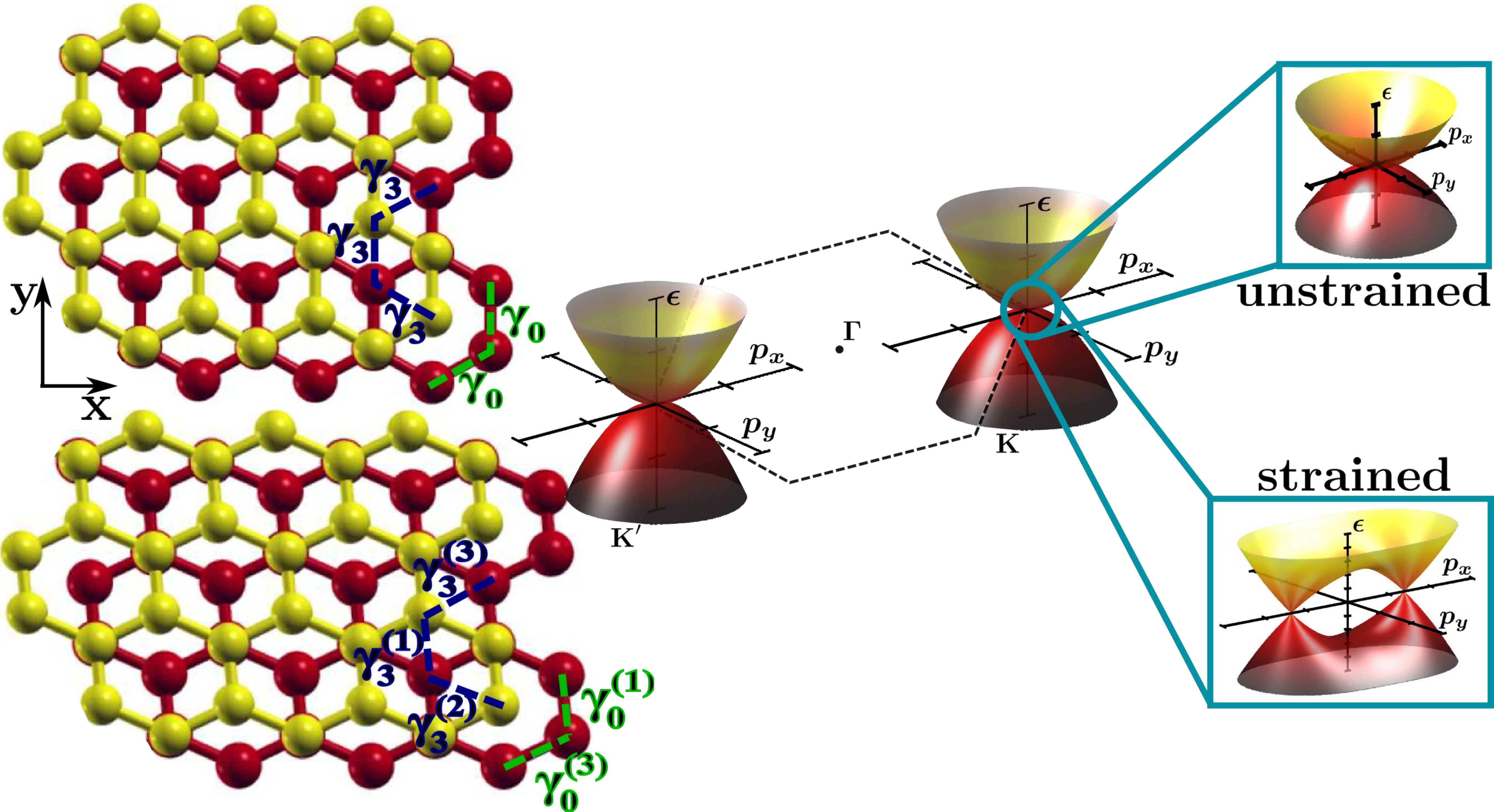}
\caption{\label{fig1} (Color online) Left: Top view of an unperturbed (top panel) and a strained (bottom panel) bilayer graphene (BLG) crystal. The top and bottom layers are shown in yellow and red, respectively. Strain modifies the intra-layer nearest neighbors coupling $\gamma_0$, as well as the inter-layer coupling $\gamma_3$ between atoms at the center of the other layer's hexagons. Right: Electronic band structure in the vicinity of the Brillouin zone corners \textbf{K} and \textbf{K$^\prime$}, with focus on the low-energy dispersion near the \textbf{K} point for unperturbed and strained BLG.}
\end{figure}

Bilayer graphene (BLG),\cite{ed1,ed3} a crystal consisting of two graphene monolayers arranged according to Bernal stacking,\cite{bernal} is a material with versatile properties. In contrast to monolayer graphene, where the linear dispersion (Dirac cones) near each corner of the Brillouin zone (\textbf{K} and \textbf{K$^\prime$} points) is very difficult to alter, the low-energy band structure of BLG can be qualitatively modified by relatively weak external perturbations. For example, a transverse electric field opens a mini-gap in the BLG spectrum.\cite{ed1,ohta,ed4,castro-neto1} Also, it has been shown that a relatively small uniaxial strain (of only a few percent) leads to a change in the topology of the low-energy dispersion, which then exhibits two Dirac mini-cones near each corner of the Brillouin zone  (see Fig. \ref{fig1}).\cite{marcin,mariani,son,lemonik}
Both in the conduction and valence bands, these cones are connected by a saddle point at which the Fermi lines reconnect, a configuration which is known as a Lifshitz transition (LiTr).\cite{lifshitz,abrikosov}  (In contrast, in a monolayer, homogeneous strain only results in a small shift of the Dirac cones away from the corners of the Brillouin zone, without any qualitative change of the linear dispersion or the chiral properties of the electrons.\cite{castro-neto})

In this paper, we study transport characteristics of an ideally clean homogeneously strained BLG crystal, aiming to find features in the temperature and chemical potential dependence of its two-terminal conductance that would reflect the presence of the saddle point in the dispersion relation.  We consider a short and infinitely wide strained BLG strip (the only geometry where strain in a two-terminal device would be homogeneous\cite{marcin1}), adjacent to BLG regions suspended over metallic contacts. The information encoded in the two-terminal conductance of such a device of finite length is complementary to what is manifested by the sheet conductivity of an infinite flake discussed in Ref.~[\onlinecite{cserti}]. Since contacts with metals heavily dope graphene, we model the BLG terminals with a high (\emph{e.g.}, n-type) density of carriers, whereas the strip in the middle is considered to be at a low density of carriers (either of n- or p-type, which can be controlled by an external gate). We choose the amount of strain in the structure such that it induces a LiTr at the energy of about $\pm5$ meV, measured from the charge neutrality point. According to Ref.~[\onlinecite{marcin}], such an effect on the bands can be generated by about $\sim1\%$ of uniaxial strain. Note that in suspended graphene structures\cite{weitz,velasco,mayorov} strain of such size may be inflicted involuntarily, either by processing and annealing of the flake, or by displacements of contacts due to the different contractions upon cooling of the substrate and the supporting metallic electrodes.
Our findings show that the dependence of the conductance $G(\mu,T)$ on the chemical potential and temperature  does indeed reflect the spectral reconstruction by strain, in the form of a conductance anomaly: a non-monotonic dependence of $G(\mu,T)$ on both parameters, $\mu$ and $T$. This behavior is characteristic for the regime where the chemical potential is close to the saddle point in the electron/hole spectrum on one of the sides of an $n$-$p$ junction, or in the middle of an $n$-$p$-$n$ device. The conductance anomaly is sensitive to the crystallographic orientation of the sample geometry, which determines the principal axis of the strain tensor. These results are described in detail in Sections \ref{sec:np} and \ref{sec:npn}, with their graphic representation shown in Figs.~\ref{fig3} and \ref{fig4}.
The calculations are based on the Landauer-B{\"u}ttiker approach,\cite{buttiker} with transmission probabilities obtained in the transfer matrix method.\cite{nazarov,henning,peeters}
Section \ref{sec:modes} introduces the model for a strained bilayer graphene device and identifies the propagating and evanescent modes required for these calculations.

\section{\label{sec:modes}Electron dispersion and propagating modes in strained BLG}
In this section, we identify the energy dispersion and transport modes in homogeneously strained BLG regions. These results are used in the subsequent sections to study the transport in devices made out of several such regions
($n$-$p$ and $n$-$n'$ junctions in Sec.~\ref{sec:np}, $n$-$p$-$n$ and $n$-$n'$-n junctions in Sec.~\ref{sec:npn}).

The lattice structure and parametrization of the minimal relevant tight-binding model for electrons in strained BLG\cite{ed1,marcin} are illustrated in the left panel of Fig.~\ref{fig1}. The stacked layers have every $A$ site within each layer surrounded by three $B$ sites and $\it{vice}$ $\it{versa}$, with intralayer coupling $\gamma_0\sim3$eV; $A_2$ sites are on top of $B_1$ sites, with interlayer coupling $\gamma_1\sim0.4$eV, while $A_1$/$B_2$ sites sit over/under the hexagons in the other layer and are coupled by `skew' hopping energy $\gamma_3\sim0.3$eV. The low-energy electronic states reside on the sites $A_1$ and $B_2$, while the sites $A_2$ and $B_1$ support states in split bands which do not contribute to low-energy transport. For unstrained BLG, the low-energy states near each corner of the Brillouin zone form two approximately parabolic bands, a valence band and a conductance band, which touch each other at the \textbf{K} or \textbf{K$^\prime$} point, as shown in the top right panel of Fig.~\ref{fig1}.

Uniaxial strain changes the intralayer and interlayer hopping integrals $\gamma_0$ and $\gamma_3$ by making them direction dependent, as shown in the bottom left panel of Fig.~\ref{fig1}. Neglecting trigonal warping for large enough strain, the corresponding low-energy dispersion near a given corner of the Brillouin zone is described by the effective Hamiltonian\cite{marcin,mariani,son}
\begin{equation}
  \mathcal{H}  =
       \left( \begin{array}{cc}
                        V(x)     &   - \frac{1}{2 m}\left( \pi^{\dagger}\right)^2 + w e^{-2 i \phi}  \\
                - \frac{1}{2 m}\left( \pi\right)^2 + w e^{2 i \phi}   &   V(x)
                             \end{array} \right).
\label{eqn-ham1}
\end{equation}
Here $m\approx0.035m_e$ is the effective mass, $\pi= p_x+i p_y$ parametrizes the in-plane momentum relative to the \textbf{K} or  \textbf{K$^\prime$} point, and $w e^{-2 i \phi}$ accounts for the change of the couplings due to the strain, where $\phi$ is the angle between the principal axis of the strain tensor and the crystallographic direction of the crystal. Using the tight binding model for BLG, one finds\cite{marcin,mariani,son} that $w=(3/4)(\eta_3-\eta_0)\gamma_3(\delta-\delta^{\prime})$, with $\eta_{0,3}=d \ln \gamma_{0,3}/d \ln r_{AB}$ where $r_{AB}$ is the distance between carbon sites, while $\delta$ and $\delta^{\prime}$ are the two principal values of the strain tensor.

Near each corner of the Brillouin zone, the low-energy dispersion relation obtained from Eq.~\eqref{eqn-ham1} exhibits two Dirac mini-cones, which are separated from the parabolic spectrum at high energies ($w\ll|\epsilon|<\gamma_1/2$) by a saddle point at $\epsilon=\pm w$ (see bottom right panel of Fig.~\ref{fig1}). For energies $|\epsilon|<w$ between the saddle points, each mini-cone results in a disconnected, approximately circular Fermi line. At the saddle point, the lines connect pairwise in a LiTr, and beyond the LiTr there is only a single Fermi line encircling the \textbf{K} or  \textbf{K$^\prime$} point. Relative to these corner points, the strain-induced Dirac points are positioned in the momentum plane at
\begin{equation}
{\bf p}_0=p_0\left( \cos\phi, \sin\phi\right), \text{ } p_0=\pm \sqrt{2 m w}.
\label{eqn-p0}
\end{equation}
After expanding $\mathcal{H}$ in Eq.~\eqref{eqn-ham1} in momentum ${\bf p}-{\bf p}_0$ around these Dirac points (and keeping only linear terms), we find that each is characterized by a Dirac velocity $v^*=p_0/(2 m)$.

\begin{figure}[t]
\includegraphics[width=.8\columnwidth]{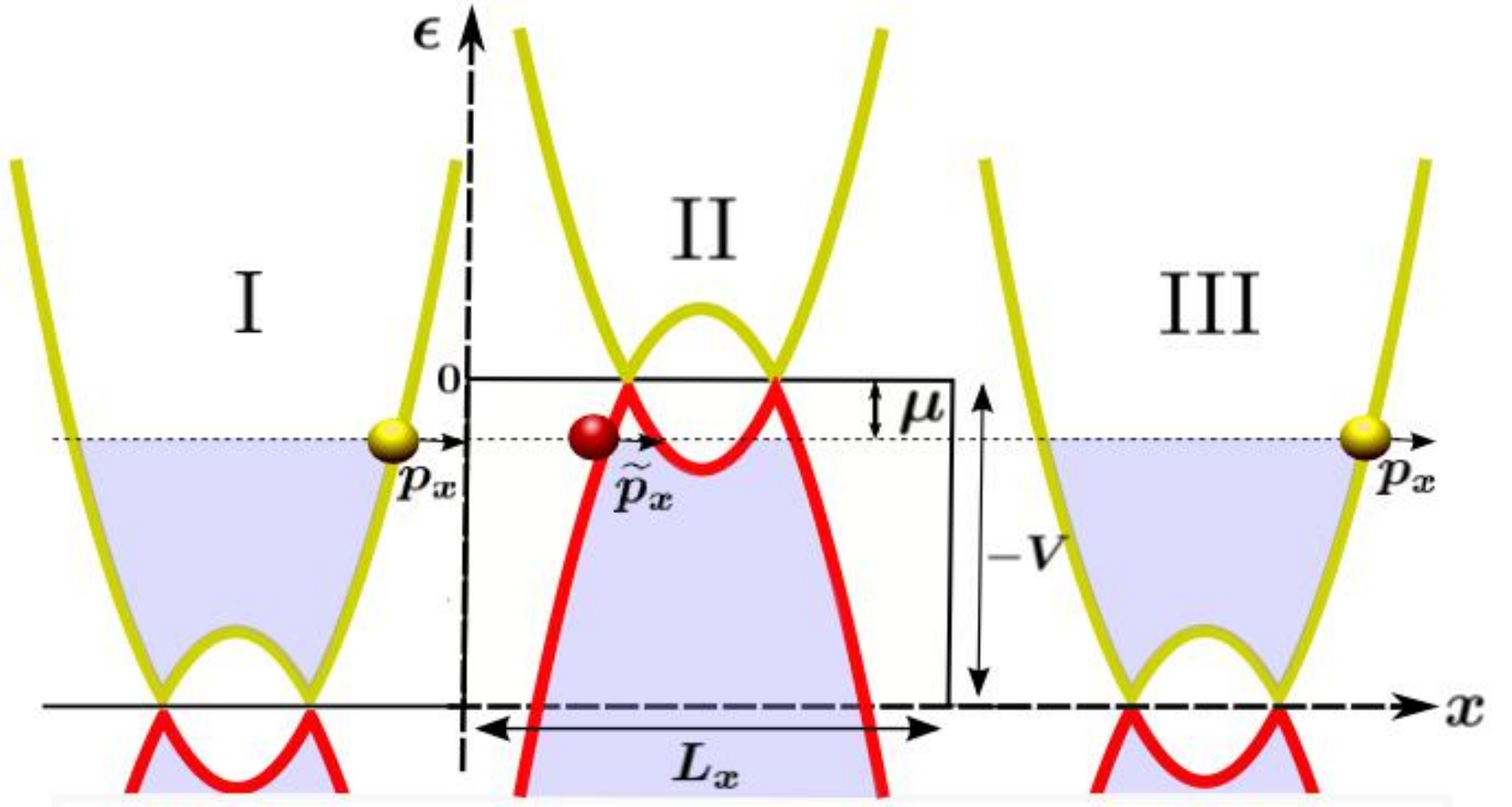}
\caption{\label{fig2} (Color online) Schematic representation of a suspended BLG device with strain axis oriented along $x$-direction (as defined in Fig.~\ref{fig1}). The sketch illustrates the example of an n-p-n configuration of such a device ($\mu<0$). In the highly doped contact regions the Fermi level~(dotted line) lies high up in the conduction band (yellow), where the dispersion is parabolic. In the central region the Fermi level lies in  the valence band (red),
and is close to the charge neutrality point, where the dispersion is modified due to the two Dirac mini-cones and the saddle point associated with the Lifshitz transition. Shading indicates occupied states.}
\end{figure}

In the following, we study how this strain-induced change in the topology of the electronic bands affects the transport properties of a device made of a strained flake of BLG, where a narrow and wide strip of width $L_y\gg L_x$ sits between two highly doped BLG regions suspended over two metallic contacts.
The band alignment in such a device is sketched in Fig.~\ref{fig2}. Metallic electrodes heavily dope BLG in the vicinity of the contacts, thus determining two leads (regions $\mathrm{I}$ and $\mathrm{III}$) with a high carrier density. An external electrostatic gate controls the doping, and, thus, the chemical potential $\mu$ of the electrons in the middle part of the flake (region $\rm{II}$), which we consider to be close to the neutrality point. In our model this doping profile is taken into account by potential steps at the sample edges
\begin{equation}
 V(x) = \left\{
  \begin{array}{l l}
    0  &  \quad \text{if} \quad 0\leq x \leq L_x, \\
    -V_0 &  \quad \text{otherwise},
  \end{array} \right. \quad \text{$\frac{\gamma_1}{2}\gg V_0\gg w$},\text{ $\mu$}. \nonumber
\end{equation}
In the remainder of this section, we identify the transport modes in the various regions of the system.

The stated conditions make the energy dispersion in the leads approximately parabolic, $\epsilon\approx p^2/(2 m)-V_0$, and the plane-wave states the same as chiral states in unstrained BLG,\cite{ed1} with a very little effect of the strain. For given incidence angle $\theta$ of an incoming electron in the contact, we parametrize their transverse momentum along the step as $p_y=\sqrt{2 m|V_0+\epsilon|}\sin(\theta)$, and use the longitudinal component
\begin{equation}
p_{x}^{l n}=l\text{ }\sqrt{n 2 m|V_0+\epsilon|-p_y^2},\quad l=\pm,
\label{eqn-mom-no}
\end{equation}
to characterize propagating modes~($n=+$, real momentum) and evanescent modes~($n=-$, complex momentum). For propagating modes in the conduction band of the leads, as considered here,  the group velocity is directed parallel to the momentum, and thus the index $l$ coincides with the propagation direction along the $x$ axis, \emph{i.e.}, $l=+$ denotes a state propagating to the right.

\begin{figure*}[t!]
\includegraphics[width=1.78\columnwidth]{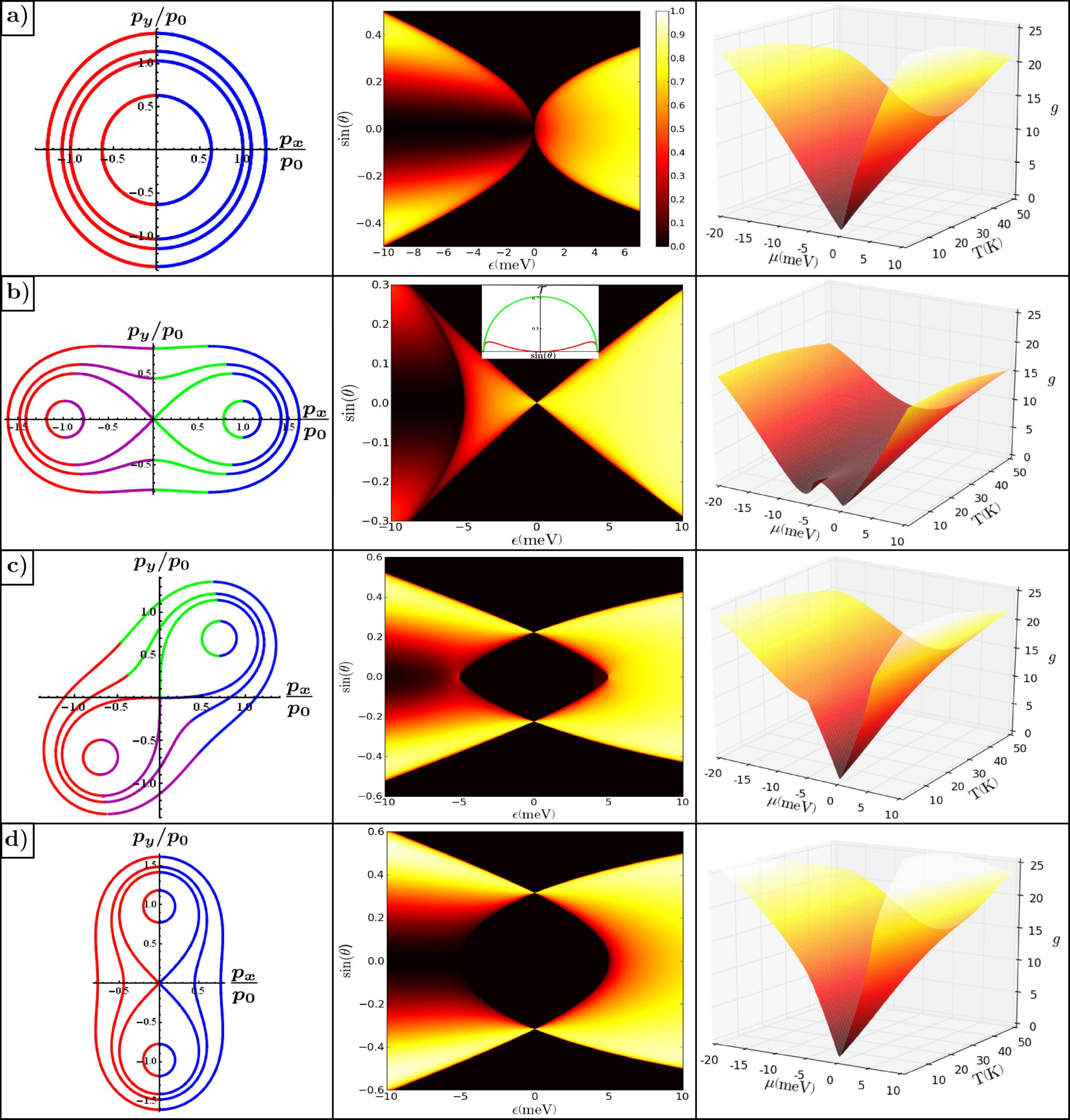}
\caption{\label{fig3} (Color online) Left: isoenergetic lines at $\epsilon=-2$, $-5$, $-6$, and $-8$ meV for strained bilayer graphene, with $w=5$ meV. Center: transmission probability $\mathcal{T}(\epsilon, \theta)$ across a single potential step (n-p or n-n$'$ junction), from a highly doped region to a barely doped region, as a function of energy and incidence angle of incoming electrons. Right: linear response conductance of the junction as a function of chemical potential $\mu$ and temperature $T$. Results are shown for unstrained bilayer graphene (a), as well as uniaxially strained bilayer graphene for various orientations of the strain axis with respect to the crystallographic axis $x$ in Fig.~\ref{fig1}:~$\phi=0$~(b), $\phi=\pi/4$~(c) and $\phi=\pi/2$~(d).}
\end{figure*}

We now turn to the modes in the weakly doped region $\mathrm{II}$.
The left panels in Fig.~\ref{fig3} show isoenergetic lines for electrons in the valence band at low energies for unperturbed [Fig.~\ref{fig3}(a)] as well as strained BLG [Figs.~\ref{fig3}(b)-\ref{fig3}(d)], with $w=5$ meV and for several orientations of the strain principal axis. States corresponding to plane waves moving to the right are indicated by red and green, and to the left by purple and blue. (Note that for some of these modes, the group velocity is directed opposite to their momentum.) These isoenergetic lines reflect that the low-energy dispersion relation in strained graphene is determined by the modified condition
\begin{align}
\epsilon^2 &= \frac{1}{4 m^2} \left(\widetilde{p}_x^2+p_y^2 \right)^2
                               -\frac{1}{m} w \left(\widetilde{p}_x^2-p_y^2 \right) \cos(2\phi) \nonumber \\
                           &-\frac{1}{m}2 w \widetilde{p}_x p_y \sin(2 \phi) +w^2
\label{eqn-disp}
\end{align}
(here and in the following, an overscript tilde denotes quantities specific for region $\mathrm{II}$; energy $\epsilon$ and transverse momentum $p_y$ are conserved for elastic scattering at a straight interface).
For given values of energy and transverse momentum,
this equation may have four, two, or no real solutions $\widetilde{p}_x$, where the latter situation arises at any given fixed energy beyond a critical value $|p_y|=p_{y,c}$, which depends on the orientation of the applied strain.
We denote the corresponding propagation direction in region I by $\theta_c$, $p_{y,c}=\sqrt{2 m|V_0+\epsilon|}\sin(\theta_c)$, which signifies the critical angle beyond which electrons from the lead only couple into evanescent modes, which do not contribute toward transport. As such, restricting the analysis to the range of angles $(-\theta_c,\theta_c)$ is enough to capture all essential transport features. Below the critical value and for large values of $|\epsilon|$, there are always two real and two complex solutions, while for small $|\epsilon|$, there are two or four real solutions, which depend on the propagation direction and on the orientation of the applied strain, as we now discuss in detail.

For the unstrained case [neglecting $w$ in Eq.~\eqref{eqn-disp}], the parameters in region $\mathrm{II}$ are given
[in analogy to Eq.~\eqref{eqn-mom-no}] by
\begin{equation}
\widetilde{p}_{x}^{l n}=l\text{ }\mathrm{sign}(\epsilon)\sqrt{n 2 m|\epsilon|-p_y^2}.
\label{eqn-mom-no2}
\end{equation}
Here, $\widetilde{p}_x^{++}$ ($\widetilde{p}_x^{-+}$) is real and corresponds to right-moving~(left-moving) plane waves, while $\widetilde{p}_x^{--}$ ($\widetilde{p}_x^{+-}$) corresponds to evanescent waves decaying to the right (left). [The factor $\mathrm{sign}(\epsilon)$ accounts for the fact that in the valence band, the group velocity is directed opposite to the momentum.]

For strained BLG with strain orientation $\phi=0$, we find from Eq.~\eqref{eqn-disp} that
\begin{equation}
\widetilde{p}_{x}^{ln}=l\text{ }\mathrm{sign}(\epsilon) \sqrt{n \sqrt{4 m^2 \epsilon^2-8 m w p_y^2}-p_y^2 + 2 m w}, \label{eqn-mom-0}
\end{equation}
where $n,l=\pm$.
The left panel of Fig.~\ref{fig3}(b) shows examples of several isoenergetic lines, with strain-induced Dirac points on the axis $p_x$ in the momentum space. By inspecting Eq.~\eqref{eqn-mom-0}, one notices that for $|\epsilon|<w$ and $p_y \leq \sqrt{m \epsilon^2/(2 w)}$
[angles where $\sin(\theta) \leq \sqrt{\epsilon^2/(4 w |V_0+\epsilon|)}$],
all four momenta are real [$\widetilde{p}_x^{++}$ (red), $\widetilde{p}_x^{--}$ (green), $\widetilde{p}_x^{+-}$ (purple), $\widetilde{p}_x^{-+}$ (blue)] and the Fermi line is split into two pockets. When $\epsilon$ is slightly below (above) the LiTr in the valence (conduction) band, $|\epsilon|>w$, the Fermi line is continuous but deformed. For small values of $|p_y|$, Eq.~\eqref{eqn-disp} then gives two real solutions ($\widetilde{p}_x^{++}$, $\widetilde{p}_x^{-+}$) and two imaginary solutions ($\widetilde{p}_x^{--}$, $\widetilde{p}_x^{+-}$), while for larger values of $|p_y|$ (just below the critical value $p_{y,c}$)
there are four real solutions.

Figure~\ref{fig3}(c) illustrates the propagating modes for strain with orientation $\phi=\pi/4$, where the momenta were found numerically
from Eq.~\eqref{eqn-disp}. The four colors distinguish right-moving plane waves ($\widetilde{p}_x^{++}$ red, $\widetilde{p}_x^{--}$ green) and left-moving plane waves ($\widetilde{p}_x^{+-}$ purple, $\widetilde{p}_x^{-+}$ blue). We now find at most two real solutions for fixed energy and transverse momentum. Above the LiTr in the valence band, there is a range of transverse momenta around $p_y=0$ (normal incidence from the leads) in which there are no propagating modes in region II.

For the strain axis oriented at $\phi=\pi/2$ [Fig.~\ref{fig3}(d)], the four solutions of  Eq.~\eqref{eqn-disp} are
\begin{equation}
\widetilde{p}_{x}^{ln}=l\text{ }\mathrm{sign}(\epsilon)\sqrt{n \sqrt{4 m^2 \epsilon^2+8 m w p_y^2}-p_y^2 - 2 m w}, \label{eqn-mom-90}
\end{equation}
where, as before, $n=\pm$ and $l=\pm$.
By inspecting Eq.~\eqref{eqn-mom-90}, we find that for all energies and angles below $\theta_c$, only the momenta $\widetilde{p}_x^{l+}$ are real. In Fig.~\ref{fig3}(d), the corresponding propagating waves are marked red ($\widetilde{p}_x^{++}$, right-moving) and blue ($\widetilde{p}_x^{-+}$, left-moving).
Above the LiTr in the valence band, there is again a range of transverse momenta around $p_y=0$ in which no propagating modes exist in region II.

\section{\label{sec:np}Transport across $n$-$p$ and $n$-$n^{\prime}$ junctions}
In this section, we study how the presence of strain affects the electron transmission across a single potential step, from a heavily doped region $\mathrm{I}$ to the low-density region $\mathrm{II}$. Depending on the sign of the doping, this can be an $n$-$p$ ($\mu<0$) or $n$-$n^{\prime}$ ($\mu>0$) junction. In the Landauer-B{\"u}ttiker approach,\cite{buttiker,nazarov,henning,peeters} the conductance $G$ of such a junction is determined by the energy and angular dependence of the transmission probability $\mathcal{T}(\epsilon, \theta)$ of an electron in the conduction band incident from the left to emerge in the valence (or conduction) band at the right of the interface.

\subsection{Transmission probability}

To calculate the transmission probability $\mathcal{T}(\epsilon, \theta)$, we employ the transfer matrix method.\cite{nazarov} Using separation of variables~(allowed for a straight interface), the spinor eigenstates of Eq.~\eqref{eqn-ham1} can be written as $\Phi_{\mathrm{I},\mathrm{II}}(x,y)= \Psi_{\mathrm{I},\mathrm{II}} \left( x \right) e^{i p_{y} y}$, where
\begin{align}
\Psi_{\mathrm{I}} \left( x \right)&=\sum_{l,n} \frac{a_{l n}}{\sqrt{v_{ln}}} \left( \begin{array}{c}
                                                             1 \\
                                                             \alpha_{l n}
                                                \end{array} \right) e^{i p_{x}^{l n} x}, \label{eqn-wave1}  \\
\Psi_{\mathrm{II}} \left( x \right)&=\sum_{l,n} \frac{b_{l n}}{\sqrt{\widetilde{v}_{ln}} }\left( \begin{array}{c}
                                                             1 \\
                                                             \beta_{l n}
                                                \end{array} \right) e^{i \widetilde{p}_{x}^{l n} x}, \nonumber \\
\alpha_{l n}&=\frac{-\frac{1}{2m}\left(p_{x}^{l n}+i p_y\right)^2}{V_0+\epsilon},\nonumber \\
\beta_{ln}&= \frac{-\frac{1}{2m}\left(\widetilde{p}_{x}^{ln}+i p_y\right)^2+w e^{2 i \phi}}{\epsilon}. \nonumber
\end{align}
Here, as before, indices $\mathrm{I}$ and $\mathrm{II}$ label regions to the left and right from the potential step, $l,n=\pm$ discriminate the branches of longitudinal momentum, $a_{l n}$, $b_{l n}$ are the wave amplitudes, and $v_{l n}=\left| \partial \epsilon/ \partial p_{x}^{ln} \right|$, $\widetilde{v}_{l n}=\left|\partial \epsilon/\partial \widetilde{p}_{x}^{ln}\right|$ are the longitudinal components of the group velocity. [Note that in the conduction band ($\epsilon>0$) and in the valence band ($\epsilon<0$), the group velocities for fixed electron momentum are oppositely directed.]

The transfer matrix $M_1$ relates the amplitudes $a_{ln}$ and $b_{ln}$ on the two sides from the interface according to
\begin{align}
 \left( \begin{array}{c}
                  b_{++}\\
                  b_{-+}\\
                  b_{--}\\
                  b_{+-}
                       \end{array} \right)&=
 M_1\left( \begin{array}{c}
                  a_{++}\\
                  a_{-+}\\
                  a_{--}\\
                  a_{+-}
                       \end{array} \right). \label{eqn-transfer-np1}
\end{align}
To build this matrix, we employ the continuity of the electron wave function $\Psi_\mathrm{I} \left( 0 \right)=\Psi_\mathrm{II} \left( 0 \right)$ and its derivative $\Psi_\mathrm{I}^{\prime} \left( 0 \right)=\Psi_\mathrm{II}^{\prime} \left( 0 \right)$ at the potential step. In this way, we find that the transfer matrix takes the form
\begin{equation}
M_{1}=B^{-1} A, \label{eqn-transfer1}
\end{equation}
with

\begin{align}
A =&\left( \begin{array}{cccc}
                  1     &    1  & 1 & 1\\
                  \alpha_{++}& \alpha_{-+}& \alpha_{--} & \alpha_{+-}\\
                  p_x^{++}& p_x^{-+}& p_x^{--} & p_x^{+-}\\
                  \alpha_{++}p_x^{++} & \alpha_{-+}p_x^{-+} & \alpha_{--}p_x^{--}&  \alpha_{+-}p_x^{+-}
                       \end{array} \right)\nonumber \\
& \times \left(\begin{array}{cccc}
                  \frac{1}{\sqrt{v_{++}}} & 0 &0 &0\\
                  0& \frac{1}{\sqrt{v_{-+}}} &0 &0\\
                  0& 0& \frac{1}{\sqrt{v_{--}}} &0\\
                  0& 0& 0& \frac{1}{\sqrt{v_{+-}}}
                       \end{array} \right), \nonumber \\
B =&\left(\begin{array}{cccc}
                  1     &    1  & 1 & 1\\
                  \beta_{++}& \beta_{-+}& \beta_{--} & \beta_{+-}\\
                  \widetilde{p}_x^{++}& \widetilde{p}_x^{-+}& \widetilde{p}_x^{--} & \widetilde{p}_x^{+-}\\
                  \beta_{++}\widetilde{p}_x^{++} & \beta_{-+}\widetilde{p}_x^{-+} & \beta_{--}\widetilde{p}_x^{--}&  \beta_{+-}\widetilde{p}_x^{+-}
                       \end{array} \right) \nonumber \\
&\times \left( \begin{array}{cccc}
                  \frac{1}{\sqrt{\widetilde{v}_{++}}} &0 &0 &0\\
                  0& \frac{1}{\sqrt{\widetilde{v}_{-+}}} &0 &0\\
                  0& 0& \frac{1}{\sqrt{\widetilde{v}_{--}}} &0\\
                  0& 0& 0& \frac{1}{\sqrt{\widetilde{v}_{+-}}}
                       \end{array} \right). \nonumber
\end{align}

In what follows, we characterize waves by their corresponding momentum and amplitude. In the contact, we assume that there are right-moving propagating waves ($p_x^{++}$, $a_{++}=1$) which can be transmitted into region $\mathrm{II}$ or reflected by the potential step $V_0$ back into region $\mathrm{I}$. Reflected waves become left-moving propagating waves ($p_x^{-+}$, $a_{-+}\neq0$) and evanescent waves decaying to the left ($p_x^{--}$, $a_{--}\neq0$). From this, Eq.~\eqref{eqn-transfer-np1} becomes
\begin{equation}
 \left( \begin{array}{c}
                  b_{++}\\
                  b_{-+}\\
                  b_{--}\\
                  b_{+-}
                       \end{array} \right)=
 M_1\left( \begin{array}{c}
                  1\\
                  a_{-+}\\
                  a_{--}\\
                  0
                       \end{array} \right),
\label{eqn-transfer-np2}
\end{equation}
and the transmission coefficient can be found using
\begin{equation}
\mathcal{T}=1-|a_{-+}|^2. \label{eqn-transmission}
\end{equation}
This definition is the most convenient for the problem studied here since there is only one left-propagating mode in region $\mathrm{I}$, whereas there are parametric regimes in which two different right-propagating waves exist in region $\mathrm{II}$. In the following, we discuss how this scheme is applied in the four characteristic cases illustrated in Fig.~\ref{fig3}: (a)~unstrained BLG, and (b)-(d)~BLG with various angles between the principal axis of uniaxial strain and the crystallographic direction $x$ in Fig.~\ref{fig1}.

For the unstrained case, using the plane-wave parameters determined in Eq.~\eqref{eqn-mom-no2}, Eq.~\eqref{eqn-transfer-np2} becomes
\begin{equation}
 \left( \begin{array}{c}
                  b_{++}\\
                  0\\
                  0\\
                  b_{+-}
                       \end{array} \right)=
 M_1\left( \begin{array}{c}
                  1\\
                  a_{-+}\\
                  a_{--}\\
                  0
                       \end{array} \right).
\label{eqn-trans-cond}
\end{equation}
Solving for the wave amplitude $a_{-+}$ numerically and then using Eq.~\eqref{eqn-transmission}, we obtain the transmission probability shown in the middle panel of Fig.~\ref{fig3}(a). This reproduces the $\epsilon\rightarrow -\epsilon$ asymmetry for transmission of normally incident electrons ($\theta=0$), with vanishing $\mathcal{T}(\epsilon,0)=0$ for $\epsilon<0$ but finite $\mathcal{T}(\epsilon,0)$ for $\epsilon>0$, found in earlier studies of BLG junctions and the Klein paradox\cite{katsnelson,poole,tudorovsky} (as opposed to the perfect transmission for $\theta=0$ in monolayer graphene junctions\cite{katsnelson,tudorovsky}).
This asymmetry can be attributed to the different chirality of charge carriers in the conduction and valence bands.

For the strain axis oriented at $\phi=0$, using the plane-wave parameters determined in Eq.~\eqref{eqn-mom-0} and solving for the amplitude $a_{-+}$ in the set of linear equations in Eq.~\eqref{eqn-transfer-np2}, we numerically obtain the transmission result plotted in the middle panel of Fig.~\ref{fig3}(b). Our result shows $\mathcal{T}(\epsilon,\theta=0)\neq 0$ at any $|\epsilon|<w$, as opposed to unstrained case in Fig.~\ref{fig3}(a).
The difference between transmission at $\theta=0$ for unstrained and strained BLG can be explained as follows. Expanding the Hamiltonian~\eqref{eqn-ham1} in momentum around the Dirac points $\pm {\bf p}_0$ of the strain-induced mini-cones [Eq.~\eqref{eqn-p0}], and keeping only linear terms, we find two Hamiltonians valid at $|\epsilon|\ll w$:
\begin{equation}
\mathcal{H}_{\pm}  \approx
        \pm v^* \left( \begin{array}{cc}
                  0     &   (e^{i \phi}\overline{\pi})^{\dagger}  \\
      e^{i \phi} \overline{\pi}   &    0
                       \end{array} \right), \text{ } \overline{\pi}=2 \left(\delta p_x+ i \delta p_y \right),
\label{eqn-ham2}
\end{equation}
where $\delta {\bf p}$ is a small deviation of the electron momentum from $\pm {\bf p}_0$, $v^*=p_0/(2 m)$ is the effective Dirac velocity, and $e^{i \phi}$ is a phase factor which determines the position of the Dirac points in the momentum plane. By solving the Schr{\"o}dinger equation for each of these Hamiltonians  and then employing Eq.~\eqref{eqn-transfer-np2}, we can compute the transmission probability due to the states in each cone separately. The small insert in the middle panel of Fig.~\ref{fig3}(b) shows $\mathcal{T}(\theta)$ at $\epsilon=-0.2$ meV for the left and the right Dirac mini-cones in red and green, respectively. Transmission to the left mini-cone is zero at $\theta=0$ and increases away from normal incidence, similar to the case of parabolic dispersion. Transmission to the right mini-cone, on the other hand, exhibits a maximum at $\theta=0$ and slowly decreases for angles away from $\theta=0$, which resembles the situation for monolayer graphene. \cite{vadim} Therefore, the strain-induced mini-cones modify the chirality of the low-energy states.

For the strain axis oriented at $\phi=\pi/4$, we first verify numerically for every angle of incidence and energy which momenta correspond to plane waves moving to the right (left) and evanescent waves decaying to the right (left), respectively. Then, we use Eqs.~\eqref{eqn-transfer1} and~\eqref{eqn-transfer-np2} and solve for the amplitude $a_{-+}$ of the wave reflected back into the lead $\mathrm{I}$, taking into account all physically allowed evanescent and propagating modes in region $\mathrm{II}$. The transmission shown in the middle panel of Fig.~\ref{fig3}(c) exhibits two distinct peaks, as long as the Fermi line is split into two pockets. For a small energy range below the LiTr in the valence band, $\mathcal{T}\left(\epsilon,\theta=0\right)\neq0$, which again can be attributed to the strain-induced modification of chirality of the low-energy states. Beyond the LiTr, where the effect of strain becomes weaker and the Fermi line becomes circular, we find that $\mathcal{T} \left(\epsilon, \theta=0 \right)\rightarrow0$.

For the strain axis oriented at $\phi=\pi/2$ [with plane-wave parameters determined in Eq.~\eqref{eqn-mom-90}], for all transverse momenta and energies allowing for propagating states in region $\mathrm{II}$ the corresponding linear system of equations is again the same as in Eq.~\eqref{eqn-trans-cond}.
By solving these equations numerically, we obtain the transmission probability shown in Fig.~\ref{fig3}(d). As a function of incidence angle $\theta$, the transmission now exhibits two distinct peaks for all energies in the considered range. As in the non-strained case [Fig.~\ref{fig3}(a)], this orientation of the strain delivers $\mathcal{T}(\epsilon<0,\theta=0)=0$.

Irrespective of the modifications of chirality, in all four cases there is a marked difference in the transmission strength for $\epsilon>0$ and $\epsilon<0$. For $\epsilon<0$, the interface is an $n$-$p$ junction and an electron incoming in the conduction band of the lead (region $\mathrm{I}$) emerges in the valence band at the right from interface (region $\mathrm{II}$). For $\epsilon>0$, the electron stays in the conduction band both at the left and right from the interface, which is a better transmitting $n$-$n^{\prime}$ junction.

\subsection{Conductance}
\begin{figure*}[t!]
\includegraphics[width=1.75\columnwidth]{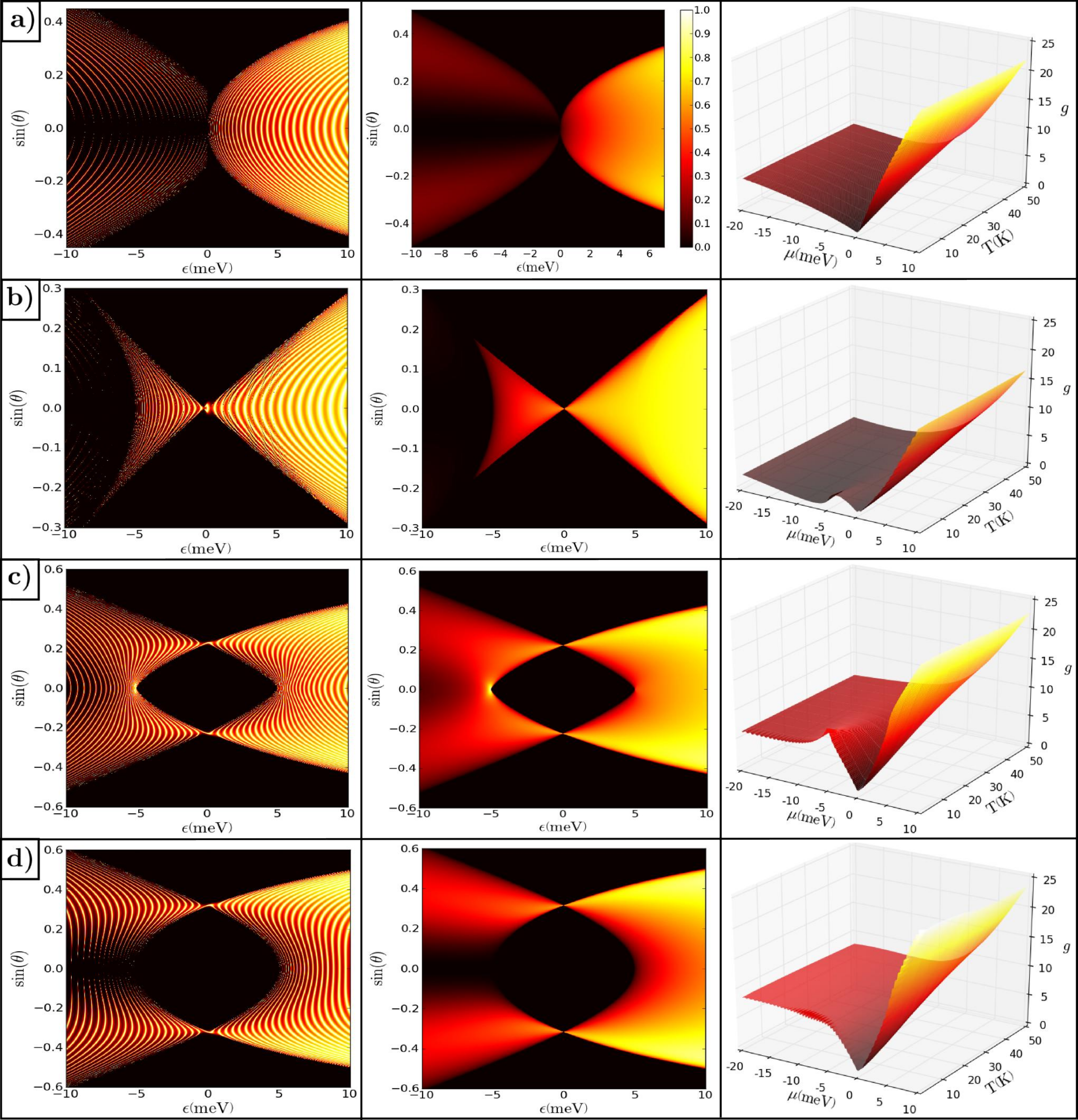}
\caption{\label{fig4} (Color online) Transmission coefficient and conductance of n-p-n and n-n$^{\prime}$-n  junctions with non-strained bilayer graphene~(a), as well as strained bilayer graphene with the uniaxial strain axis at an angle for $\phi=0$~(b), $\phi=\pi/4$~(c) and $\phi=\pi/2$~(d) from crystallographic axis $x$. Left: transmission probability $\mathcal{T}(\epsilon, \theta)$ obtained in an exact calculation. Center: transmission probability obtained by averaging over fast oscillations after the contribution of evanescent waves is neglected. Right: linear response conductance as a function of chemical potential and temperature. All calculations are performed for experimentally accessible values $w=5$ meV, $V_0=50$ meV, and $L_x=1$ $\mu$m.}
\end{figure*}
Based on the above results for the transmission probability, we employ the Landauer-B{\"u}ttiker formalism\cite{buttiker} to calculate the conductance of the $n$-$p$ or $n$-$n'$ junction. Taking into account two valleys and two spins, as well as integrating over the angle of incidence and electron energy (as determined by the Fermi distribution with finite temperature $T$), we arrive at the junction conductance,
\begin{align}
G\left(\mu, T\right)&= \frac{4 e^2}{h} \frac{L_y}
{\lambda_F} g \label{eqn-cond2} ,  \\
     g &=\frac{1}{2 \pi}\frac{1}{4 k_B T} \int_{- \infty}^{\infty}d\epsilon \frac{\sqrt{\frac{2 m \lambda_F^2}{\hbar^2}|\epsilon + V_0|} }{\cosh^2 \left( \frac{\epsilon - \mu}{2 k_B T}\right)} \nonumber\\
     &\times \int_{-\pi/2}^{\pi/2}  \mathcal{T} \left(\epsilon ,\theta\right) \cos(\theta) d\theta \ . \nonumber
\end{align}
Here,  $\lambda_{F}\approx2 \pi \hbar \sqrt{1/2m V_0}$ is the Fermi wavelength in lead $\mathrm{I}$, $k_B$ is the Boltzmann constant, and $\mathcal{T}(\epsilon, \theta)$ is the transmission at fixed energy and angle of incidence, determined above. By performing the integration numerically, we obtain the conductance as a function of chemical potential and temperature, which is shown in the right panels of Fig.~\ref{fig3}.

For unstrained BLG [Fig.~\ref{fig3}(a)], where the dispersion is parabolic, the conductance exhibits a minimum at $\mu=0$ and has an asymmetric but monotonic behavior for both $\mu<0$~($n$-$p$ junction) and $\mu>0$~($n$-$n^{\prime}$ junction). Furthermore,  for $\mu=0$, the conductance increases monotonously with temperature. In the strained cases [Figs.~\ref{fig3}(b)-\ref{fig3}(d)], the increase in conductance with temperature is still seen. However, depending on the strain orientation $\phi$, $G\left(\mu, T \right)$ can be monotonic or non-monotonic. For $\phi=0$ [Fig.~\ref{fig3}(b)], the conductance at low temperatures exhibits an anomaly: in the region $\mu<0$, there is an additional local minimum, as well as a local maximum. The local maximum is located at $\mu \approx -w$ ($\mu\approx -5$ meV for parameters used in the figure), which corresponds to the LiTr energy. For $\phi=\pi/4$ [Fig.~\ref{fig3}(c)], the conductance at low temperatures exhibits a protrusion and a shift in slope, which again occur near the LiTr in the valence band.  For $\phi=\pi/2$ [Fig.~\ref{fig3}(d)], the conductance $G\left(\mu, T\right)$ is monotonic and quite similar to that calculated for BLG with a parabolic spectrum.

All conductance plots show an asymmetry about the minimum at $\mu=0$.
As discussed for the transmission probability, this difference is determined by the chiral sublattice structure of the plane waves, which in BLG suppresses the transmission at a potential step between regions of opposite polarity.
Note that the anomalous behavior at $\mu=-w$ is specific for the $n$-$p$ junction regime of the system, and does not occur in the $n$-$n'$ junction regime, which does not exhibit an anomaly at $\mu=w$. These features allow one to single out the anomalous $T$ and $\mu$ dependence of the conductance for junctions with different orientation of the strain.

\section{\label{sec:npn}Transport across $n$-$p$-$n$ and $n$-$n^{\prime}$-$n$ junctions}
All the features found in the parametric dependencies of the transmission across a single potential step appear also in the transport properties of the two-terminal ballistic device~(with two steps) sketched in Fig.~\ref{fig2}. In particular, the $\epsilon\rightarrow -\epsilon$, and $\mu \rightarrow -\mu$ asymmetry and the anomalous temperature dependence at $\mu=-w$~(in the vicinity of the saddle point in the valence band) also persist in this ``potential barrier'' geometry, and indeed are further enhanced. In addition to those, the energy and angle dependence of the transmission coefficient acquires a resonance structure due to the interference between multiply reflected waves (Fabry-P\'{e}rot resonances). To take this into account, we compute the transmission of the device sketched in Fig.~\ref{fig2} considering both interfaces, as well as the ballistic electron propagation between the interfaces.

The transfer matrix $M_{1}=B^{-1} A$ of the first interface is given by Eq.~\eqref{eqn-transfer1}. Due to symmetry, the transfer matrix of the second interface is $M_2=M_{1}^{-1}= A^{-1} B$. The transfer matrix of the whole system ($n$-$p$-$n$ or $n$-$n^{\prime}$-$n$ junction) is then given by
\begin{equation}
\Xi = M_2 S M_1=M_1^{-1} S M_1 \ ,
\end{equation}
where
\begin{equation}
S  =
       \left( \begin{array}{cccc}
                  e^{i \widetilde{p}_{x}^{ ++} L_x}     &    0  & 0 & 0\\
                  0     &    e^{i \widetilde{p}_{x}^{-+} L_x}  & 0 & 0\\
                  0     &    0  & e^{i \widetilde{p}_{x}^{--} L_x} & 0\\
                  0     &    0  & 0 & e^{i \widetilde{p}_{x}^{+-} L_x}
                       \end{array} \right)
\label{eqn-flow}
\end{equation}
describes the ballistic electron propagation inside the ``barrier'' region $\mathrm{II}$. Note that the factors in the matrices $A$ and $B$ that normalize the plane-wave states in region II to normal flux cancel out in the matrix $\Xi$. From this, we can relate the amplitudes $a_{ln}$ of the wave function in the source lead, Eq. \eqref{eqn-wave1}, to the amplitudes $c_{ln}$ of the wave function in the drain lead,
\begin{equation}
\Psi_{\mathrm{III}} \left( x \right)=\sum_{l,n} \frac{c_{l n}}{\sqrt{v_{ln}}} \left( \begin{array}{c}
                                        1 \\
                                         \alpha_{l n}
                                                \end{array} \right) e^{i p_{x}^{l n} x}, \nonumber
\end{equation} by
\begin{equation}
 \left( \begin{array}{c}
                  c_{++}\\
                  c_{-+}\\
                  c_{--}\\
                  c_{+-}
                       \end{array} \right)=
 \Xi \left( \begin{array}{c}
                  a_{++}\\
                  a_{-+}\\
                  a_{--}\\
                  a_{+-}
                       \end{array} \right). \nonumber
\end{equation}
To determine the transmission coefficient
\begin{equation}
\mathcal{T}\left( \epsilon, \theta\right)=|c_{++}|^2,
\end{equation}
we take boundary conditions $c_{-+}=c_{--}=a_{+-}=0$, $a_{++}=1$, and find $c_{++}$ by solving the equation
\begin{equation}
 \left( \begin{array}{c}
                  c_{++}\\
                  0\\
                  0\\
                  c_{+-}
                       \end{array} \right)=
 \Xi \left( \begin{array}{c}
                  1\\
                  a_{-+}\\
                  a_{--}\\
                  0
                       \end{array} \right). \nonumber
\end{equation}

The numerically evaluated transmission probability $\mathcal{T}$ is plotted in the left panels of Fig.~\ref{fig4}, for the same range of angles and energies ($-10$ meV$<\epsilon<10$ meV, $V_0=50$ meV and $w=5$ meV) as in Fig.~\ref{fig3}. As in Refs.~[\onlinecite{katsnelson,beenakker,tudorovsky}], the presence of two reflective interfaces in a BLG device causes the appearance of resonances with high transmission. In the figure, these are seen as bright strips. The scale of the oscillations becomes finer for a longer sample length.

The right column in Fig.~\ref{fig4} shows the finite-temperature conductance in a long sample.
The interference fringes are washed out
by smearing of the Fermi step [at $k_B T\gg \hbar^2/(m L_x^2)$] and the integration over the angle. To obtain this finite-temperature conductance, one can use the exactly calculated $\mathcal{T}(\epsilon, \theta)$ and insert this into Eq.~\eqref{eqn-cond2}. Here, we describe an accurate approximation of these results, which allows one to relate the pronounced anomalies of the finite-temperature conductance to angularly smoothed transmission probabilities $\left<\mathcal{T}\right>$ (smeared over a small angle range $\delta \theta$ covering many oscillations),  shown in the middle column in Fig.~\ref{fig4}.
Conveniently, in the limit of $L_x\rightarrow \infty$, evanescent modes die off before reaching the second interface, so that in region $\mathrm{II}$ only plane waves (with real $\widetilde{p}_x$) contribute toward transmission. To eliminate the negligible contribution of evanescent waves, we first restrict the analysis to the range of angles $\Delta \theta=2 \theta_c$ where plane waves exist inside the barrier; $\theta_c$ is energy dependent and different for each orientation of applied strain. Then, we group the exponents which emerge from Eq.~\eqref{eqn-flow} into propagating and decaying waves~(where the latter have complex $\widetilde{p}_x$), and for decaying waves approximate $\tanh(|\mathrm{Im}[\widetilde{p}_x]| L_x)\rightarrow 1$ and $\cosh^{-1}(2|\mathrm{Im}[\widetilde{p}_x]| L_x) \rightarrow 0$. The conductance then follows from
\begin{align}
G\left(\mu, T\right)&= \frac{4 e^2}{h} \frac{L_y}{\lambda_F} g \ , \label{eqn-cond3} \\
 g &=\frac{1}{2 \pi}\frac{1}{4 k_B T} \int_{- \infty}^{\infty}d\epsilon \frac{\sqrt{\frac{2 m \lambda_F^2}{\hbar^2}|\epsilon + V_0|} }{\cosh^2 \left( \frac{\epsilon - \mu}{2 k_B T}\right)} \nonumber\\
     &\times \int_{-\theta_c}^{\theta_c}  \left<\mathcal{T} \left(\epsilon ,\theta_0\right)\right> \cos(\theta_0) d\theta_0.  \nonumber
\end{align}

Since the details of the analysis of $\left<\mathcal{T}\right>$ depend on the electron energy and orientation of the strain axis, we sketch the derivation separately for the corresponding characteristic parametric regimes.

Firstly, for the range of parameters for which Eq.~\eqref{eqn-disp} has only two real solutions, the described procedure leads to an expression of the form
\begin{equation}
\mathcal{T}\left( \epsilon, \theta\right) = \frac{X_1}{X_2 + X_3 \cos \left( 2 \widetilde{p}_x^{++} L_x\right) + X_4\sin \left(2 \widetilde{p}_x^{++} L_x\right)}. \nonumber
\end{equation}
Here, $X_i$ are non-oscillating functions of $p_x^{ln}$, $\widetilde{p}_x^{ln}$, $p_y$, $\epsilon$, $V_0$, $w$, and $\phi$, which are not given explicitly due to their complexity. To average $\mathcal{T}(\epsilon, \theta)$, we first expand the real momenta in terms of small deviations $\delta \theta$ in the angle, $\theta=\theta_0 +\delta \theta$, about some $-\theta_c<\theta_0<\theta_c$, such that
\begin{equation}
\widetilde{p}_x^{++}=\left. \widetilde{p}_x^{++}\right|_{\theta=\theta_0}+ \left. \delta \theta \left(\frac{\partial \widetilde{p}_x^{++}}{\partial \theta}\right) \right|_{\theta=\theta_0}. \nonumber
\end{equation}
As such,
\begin{equation}
\mathcal{T}\left( \epsilon, \theta_0\right) = \frac{X_1}{X_2 + X_3 \cos \left( \Phi+\mathcal{A}\delta \theta\right) + X_4\sin \left(\Phi+\mathcal{A}\delta \theta\right)}, \nonumber
\end{equation}
where $\Phi=2 L_x \left. \widetilde{p}_x^{++}\right|_{\theta=\theta_0}$ and $\mathcal{A}=2 L_x \left. \left(\partial \widetilde{p}_x^{++}/\partial \theta\right) \right|_{\theta=\theta_0}$. Imposing $\mathcal{A}\Delta \delta \theta=2\pi$, the average transmission over one period is
\begin{align}
\left<\mathcal{T}(\epsilon,\theta_0) \right>&= \frac{1}{2\pi} \int_{0}^{2\pi}\frac{X_1 dz}{X_2 + X_3 \cos(z) + X_4\sin(z)} \label{eqn-trans-high1}\\
&=\frac{X_1}{\sqrt{X_2^2-X_3^2-X_4^2}},  \quad \quad z=\mathcal{A}\delta \theta. \nonumber
\end{align}

Secondly, for the range of parameters where Eq.~\eqref{eqn-disp} has four real solutions, $\widetilde{p}_x^{-+}=-\widetilde{p}_x^{++}$ and $\widetilde{p}_x^{+-}=-\widetilde{p}_x^{--}$ [such as encountered in Fig.~\ref{fig3}(b)], fast oscillations in the transmission coefficient are due to combinations of $\sin(\widetilde{p}_x^{+n} L_x)$, $\sin(2\widetilde{p}_x^{+n} L_x)$, $\cos(\widetilde{p}_x^{+n} L_x)$, and $\cos(2\widetilde{p}_x^{+n} L_x)$. Expanding in terms of small deviations in angle,
\begin{equation}
\widetilde{p}_x^{+n}= \left. \widetilde{p}_x^{+n} \right|_{\theta=\theta_0}+ \left. \delta \theta \left(\frac{\partial \widetilde{p}_x^{+n}}{\partial \theta}\right) \right|_{\theta=\theta_0}, \nonumber
\end{equation}
and denoting $\Phi_n=L_x \left. \widetilde{p}_x^{+n} \right|_{\theta=\theta_0}$ and $\mathcal{A}_n=L_x \left. (\partial \widetilde{p}_x^{+n}/\partial \theta) \right|_{\theta=\theta_0}$, we find that the interference fringes are encoded in the factors $\sin(\Phi_n+\mathcal{A}_n \delta \theta)$ and $\cos(\Phi_n+\mathcal{A}_n \delta \theta)$. Inspection of the constant pre-factors reveals that $\mathcal{A}_+ \approx \mathcal{A}_-$. Neglecting the phase $\Phi_n$ and imposing $\mathcal{A}_+ \Delta \delta \theta=2\pi$, the averaged transmission over one period can then be written as
\begin{align}
&\left<\mathcal{T}\left( \epsilon, \theta_0\right) \right>= \frac{1}{2 \pi} \int_{0}^{2 \pi}\frac{\mathcal{X}(z)}{\mathcal{Y}(z)} dz,
\label{eqn-trans-high3}
\end{align}
\begin{align}
\mathcal{X}(z)=&(\bar{X}_1 \cos(z)+\bar{X}_2 \sin(z))^2, \nonumber \\
\mathcal{Y}(z)=&\bar{X}_3 + \bar{X}_4 \cos (2z) + \bar{X}_5\sin(2 z)+\bar{X}_6\cos(4z) \nonumber \\
+& \bar{X}_7 \sin(4z), \nonumber
\end{align}
and $z=\mathcal{A}_+ \delta \theta$; here $\bar{X}_i$ are non-oscillating functions of the same parameters as in the previous cases. The specific expressions are again omitted because of their complexity.

In the analytical part of the studies of the transmission problem, all functions $X_i$ and $\bar{X}_i$ have been found using the symbolic mathematical software. The results of the integrals in Eqs.~\eqref{eqn-trans-high1} and \eqref{eqn-trans-high3} are shown in the central column in Fig.~\ref{fig4}. The doping (chemical potential $\mu$) and temperature dependence of the two-terminal conductance of the device follows from Eq.~\eqref{eqn-cond3}, and coincides with a high accuracy with the one calculated using Eq.~\eqref{eqn-cond2} together with the exact values $\mathcal{T}(\epsilon,\theta)$.

The behavior of $G\left(\mu, T\right)$ in the right column of Fig.~\ref{fig4} displays all the features of the conductance of a single step in enhanced form. In particular, the conductance for $\phi=0$ [Fig.~\ref{fig4}(b)] exhibits a local maximum and a second local minimum positioned at the same chemical potentials as for a single junction. For $\phi=\pi/2$ [Fig.~\ref{fig4}(d)] the conductance is monotonic. For  $\phi=\pi/4$ [Fig.~\ref{fig4}(c)] the protrusion in the conductance of a single junction~(at the LiTr) has developed into a clear local maximum.

\section{Summary}
In this article we have shown that the linear response conductance $G(\mu, T)$ of an $n$-$p$-$n$ junction in strained bilayer graphene has a non-monotonic dependence on doping and temperature, which varies in size and form as a function of the crystallographic orientation of the principal strain axis. To understand this behavior we studied the transmission and conductance for a single interface ($n$-$p$ junction), and used the obtained results to conclude that the non-monotonic behavior is due to the modification of chirality (thus, the feature responsible for the occurrence of the Klein paradox in graphene). Uniaxial strain changes the chirality (sublattice composition) of the electronic plane-wave states in the vicinity of the saddle point (Lifshitz transition) in the low-energy electron spectrum of strained bilayer graphene, which results in the observed no-monotonicity of the linear response conductance.
\newline
\section{Acknowledgments}
We thank V.~Cheianov, A.~Geim, M.~Mucha-Kruczynski, and K.~Novoselov for useful discussions. This project was funded by EC STREP Concept Graphene, by EPSRC, and by the Royal Society.

\end{document}